\documentclass[conference]{IEEEtran}

\usepackage{cite}
\usepackage{amsmath,amssymb,amsfonts}
\usepackage{graphicx}
\usepackage{booktabs}
\usepackage{multirow}
\usepackage{array}
\usepackage{xcolor}
\usepackage[hypertexnames=false]{hyperref}
\usepackage{orcidlink}
\usepackage{url}
\usepackage{algorithm}
\usepackage{algpseudocode}

\hypersetup{
    colorlinks=true,
    linkcolor=black,
    citecolor=black,
    urlcolor=blue
}
\IEEEoverridecommandlockouts
\begin{document}

\title{GeoDistNet: An Open-Source Tool for Synthetic Distribution Network Generation}

\newcommand{\orcidYunqi}{0000-0003-1013-5497}
\newcommand{\orcidXinghuo}{0000-0001-8093-9787}
\newcommand{\orcidMahdi}{0000-0002-0517-9420}

\author{
\IEEEauthorblockN{Yunqi Wang\,\orcidlink{\orcidYunqi}, Xinghuo Yu\,\orcidlink{1, \orcidXinghuo}, and Mahdi Jalili\,\orcidlink{\orcidMahdi}}
\IEEEauthorblockA{
School of Engineering, RMIT University, Melbourne, VIC 3000, Australia 
\thanks{This work is partially supported by Australian Research Council's Discovery Program Grant DP240100830.}
}
}

\maketitle

\begin{abstract}
Distribution-level studies increasingly require feeder models that are both electrically usable and structurally representative of practical service areas. However, detailed utility feeder data are rarely accessible, while benchmark systems often fail to capture the geographic organization of real urban and suburban networks. This paper presents GeoDistNet, an open-source tool for synthetic distribution network generation from publicly available geographic information. Starting from map-derived spatial data, the proposed workflow constructs a candidate graph, synthesizes feeder-compatible radial topology through a mixed-integer formulation, assigns representative electrical parameters and loads, and exports the resulting network for power-flow analysis. A Melbourne case study shows that the generated feeder remains geographically interpretable, topologically structured, and directly usable in \texttt{pandapower} under multiple loading levels. GeoDistNet therefore provides a reproducible workflow for bridging publicly accessible GIS data and simulation-ready distribution feeder models when detailed utility networks are unavailable.
\end{abstract}


\section{Introduction}

Distribution systems are becoming a central focus in modern power system research due to the rapid growth of distributed energy resources (DERs), electric vehicles, flexible demand, and active network operation. As a result, a broad range of studies including hosting-capacity assessment, voltage regulation, service restoration, network reinforcement, and flexibility coordination, now depend on feeder models that are not only electrically analyzable, but also structurally representative of practical networks. In many of these applications, the realism of feeder topology, route layout, and load distribution directly affects the credibility of the resulting conclusions.

Despite this need, access to real utility feeder data remains highly restricted. Detailed distribution network models are typically unavailable outside utilities because of confidentiality, security, and commercial constraints. Consequently, many studies continue to rely on a limited number of benchmark feeders, such as the classical 33-bus test system proposed by reference \cite{Baran1989}. These benchmark systems remain indispensable for algorithm development and comparative assessment. More recently, efforts have also been made to improve their realism for active distribution system studies, for example through the enhanced IEEE 33-bus benchmark in \cite{Dolatabadi2020}. Nevertheless, benchmark feeders remain small in number and limited in regional diversity. They are useful as reference systems, but they do not generally preserve the geographic structure, corridor layout, or spatial heterogeneity of real urban and suburban service areas.

This limitation has been explicitly recognized in the literature. Postigo Marcos \emph{et al.} reviewed distribution test feeders used in the United States and emphasized the need for synthetic representative networks that extend beyond a few standard benchmark cases \cite{Postigo2017}. More broadly, the authors in \cite{Birchfield2017} established structural metrics as validation criteria for synthetic grids, showing that realistic network studies require synthetic models to be evaluated against topological characteristics rather than treated as arbitrary graph instances . These observations have motivated growing interest in synthetic grid construction as a practical alternative when utility-grade data are unavailable.

At the distribution level, synthetic network generation has advanced considerably in recent years. Mateo \emph{et al.} demonstrated that large-scale synthetic U.S. electric distribution system models can be constructed with realistic structural and engineering characteristics \cite{Mateo2020}. This line of work has shown that synthetic distribution models can serve as a practical bridge between idealized test systems and inaccessible utility feeders. At the same time, it also makes clear that feeder synthesis is not simply a graph-generation problem. A useful distribution network model must satisfy multiple requirements simultaneously: it should reflect plausible spatial organization, remain electrically interpretable, and support the simulation tasks for which it is intended.

Existing approaches to synthetic feeder generation can be broadly grouped into three categories. The first adapts existing benchmark feeders by modifying topology, scaling load levels, or adjusting parameters to suit a given study. The second follows statistical or engineering-template approaches, in which feeders are generated to match target topological or asset-level distributions. The third increasingly explores data-driven or graph-learning approaches. For example, Liang \emph{et al.} proposed FeederGAN, which generates synthetic feeders through deep graph adversarial learning and reproduces feeder topology and component patterns from existing feeder datasets \cite{Liang2021}. These studies confirm that synthetic feeder generation is both necessary and technically viable. However, many such approaches still depend on proprietary feeder inventories, pre-existing utility models, or statistical training data that are not themselves openly available.

A related line of work considers geographically informed feeder construction. This direction is especially relevant because practical distribution feeders are strongly shaped by road corridors, land development, and spatial demand layout. Saha \emph{et al.} proposed a framework for generating synthetic distribution feeders from OpenStreetMap, demonstrating the potential of map-derived topology as a basis for feeder construction \cite{Saha2019}. Trpovski \emph{et al.} similarly explored synthetic distribution grid generation in a planning-oriented setting, highlighting the importance of geographic and engineering consistency in feeder synthesis \cite{Trpovski2018}. These studies indicate that publicly accessible geographic information can provide a useful basis for synthetic feeder construction, particularly when confidential utility models are unavailable.

Even so, a practical gap remains between publicly available map data and simulation-ready electrical feeder models. A road map is not a distribution network. It contains geometric and spatial information, but it does not directly specify feeder radiality, source location, electrical parameters, or demand allocation. Map-derived graphs are often too dense, too detailed, or too geometrically unconstrained to be used directly in feeder simulation. As a result, a formal construction layer is required to translate raw geographic information into a connected, radial, and electrically usable network representation. This translation is nontrivial: it must preserve dominant spatial structure while removing unnecessary geometric detail, and it must do so in a way that is compatible with downstream electrical analysis.

Another practical requirement is reproducibility. If synthetic feeder generation is to be broadly useful, the construction workflow should be transparent and implementable using accessible tools. From the electrical-analysis perspective, \texttt{pandapower} has emerged as a widely used open-source framework for modeling, analysis, and optimization of electric power systems \cite{Thurner2018}. OpenDSS has likewise remained a common platform for distribution simulation studies \cite{Dugan2011}. On the geographic-data side, OSMnx provides an effective route for acquiring and processing street-network structure from OpenStreetMap \cite{Boeing2017}. Together, these developments make it increasingly feasible to build synthetic feeder workflows that are both open and practically useful.

This paper addresses the above gap through \emph{GeoDistNet}, an open-source tool for synthetic distribution network generation from publicly accessible geographic information. Rather than learning feeder structure from confidential utility models or relying solely on statistical templates, GeoDistNet starts from map-derived spatial data, extracts a candidate graph, synthesizes feeder-compatible radial topology through a mixed-integer formulation, assigns representative electrical parameters and loads, and exports the resulting model for power-flow analysis. In this sense, the contribution of the present work does not lie in proposing a new power-flow engine. Instead, it lies in bridging real-world geographic structure and simulation-ready feeder construction within an open and reproducible workflow.


The main contributions of this paper are summarized as follows:

\begin{itemize}
    \item GeoDistNet is introduced as an open-source workflow for synthetic distribution feeder generation from publicly available geographic information, bridging map-derived spatial data and simulation-ready feeder models.
    \item We formulate the feeder synthesis stage as a mixed-integer optimization problem on a candidate graph, so that the generated network remains radial, source-connected, and geographically constrained rather than being a direct reproduction of the raw street graph.
    \item The proposed workflow further incorporates representative electrical parameter assignment, aggregated demand allocation, and direct export to \texttt{pandapower}, allowing the synthesized feeder to be validated within a standard steady-state analysis environment. A case study is conducted to demonstrate the practical usability of the proposed workflow.
\end{itemize}



\section{Feeder Synthesis Formulation}

GeoDistNet constructs a synthetic feeder by selecting a connected radial subgraph from a map-derived candidate graph. The synthesis problem is formulated at the graph level, so that geographic plausibility and feeder operability can be imposed before electrical quantities are assigned. This separation is useful because raw GIS data preserve spatial structure but do not directly define a simulation-ready feeder.

\begin{figure}[!bp]
\centering
\includegraphics[width=0.85\columnwidth]{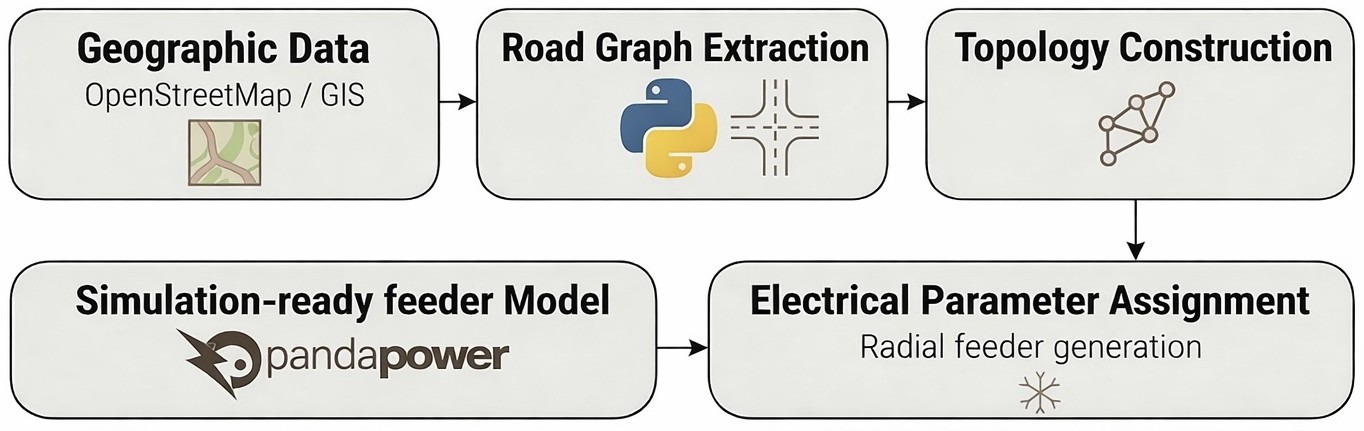}
\caption{Overall workflow of GeoDistNet. Geographic data are processed into a candidate graph, from which a radial feeder is synthesized and then converted into a simulation-ready electrical model.}
\label{fig:workflow}
\end{figure}

Fig.~\ref{fig:workflow} summarizes the overall GeoDistNet workflow. Starting from geographic data, the framework extracts a candidate graph from the road-constrained spatial environment, synthesizes a source-connected radial feeder, and then maps the selected topology to an electrical network model for subsequent simulation.

Mathematically, let the original map-derived graph be
\begin{equation}
\mathcal{G}^{\mathrm{map}} = (\mathcal{N}^{\mathrm{map}}, \mathcal{E}^{\mathrm{map}}),
\end{equation}
where $\mathcal{N}^{\mathrm{map}}$ and $\mathcal{E}^{\mathrm{map}}$ denote map-derived nodes and edges, respectively. After preprocessing and filtering, a candidate graph is obtained:
\begin{equation}
\mathcal{G}^{\mathrm{cand}} = (\mathcal{N}^{\mathrm{cand}}, \mathcal{E}^{\mathrm{cand}}).
\end{equation}

This graph retains the dominant spatial corridors and node locations relevant to feeder construction while removing unnecessary geometric detail.

The feeder synthesis step seeks a connected radial subgraph of $\mathcal{G}^{\mathrm{cand}}$. Let $y_i$ be a binary variable indicating whether candidate node $i$ is retained, and let $z_{ij}$ be a binary variable indicating whether candidate edge $(i,j)$ is activated:
\begin{equation}
y_i \in \{0,1\}, \qquad \forall i \in \mathcal{N}^{\mathrm{cand}},
\end{equation}
\begin{equation}
z_{ij} \in \{0,1\}, \qquad \forall (i,j)\in\mathcal{E}^{\mathrm{cand}}.
\end{equation}

Node and edge selections are coupled through
\begin{equation}
z_{ij} \le y_i,\qquad z_{ij} \le y_j,
\qquad \forall (i,j)\in\mathcal{E}^{\mathrm{cand}}.
\end{equation}

These constraints prevent an edge from being selected unless both of its incident nodes are retained.

Let $\mathcal{N}^{\mathrm{req}} \subseteq \mathcal{N}^{\mathrm{cand}}$ denote the set of required nodes that must remain in the synthesized feeder. Depending on the study setup, these may correspond to source-adjacent nodes, terminal anchors, or nodes associated with mandatory demand representation. For such nodes,
\begin{equation}
y_i = 1, \qquad \forall i\in\mathcal{N}^{\mathrm{req}}.
\end{equation}

A weighted objective is adopted to balance geographic fidelity, topological compactness, and electrical practicality:
\begin{equation}
\min_{y,z}
\sum_{(i,j)\in\mathcal{E}^{\mathrm{cand}}}
\left(
\alpha_{\mathrm{geo}} c_{ij}^{\mathrm{geo}}
+
\alpha_{\mathrm{top}}
+
\alpha_{\mathrm{elec}} d_{ij}
\right) z_{ij},
\label{eq:feeder_obj}
\end{equation}
where $c_{ij}^{\mathrm{geo}}$ is a geographic penalty score, $d_{ij}$ is segment length, and $\alpha_{\mathrm{geo}}, \alpha_{\mathrm{top}}, \alpha_{\mathrm{elec}}$ are nonnegative weights. The first term penalizes geographically undesirable corridors, the second discourages unnecessarily dense branching, and the third suppresses excessive total route length.

To impose radiality, the selected feeder must satisfy the tree-cardinality condition
\begin{equation}
\sum_{(i,j)\in\mathcal{E}^{\mathrm{cand}}} z_{ij}
=
\sum_{i\in\mathcal{N}^{\mathrm{cand}}} y_i - 1.
\label{eq:tree_cardinality}
\end{equation}

This condition ensures that, once connectivity is enforced, the selected subgraph is radial.

Connectivity to a fixed source node $s$ is imposed using a single-commodity flow defined on the directed arc set $\mathcal{A}^{\mathrm{cand}}$ induced from $\mathcal{E}^{\mathrm{cand}}$. Let $f_{ij}\ge 0$ denote the flow on directed arc $(i,j)\in\mathcal{A}^{\mathrm{cand}}$. For every retained non-source node,
\begin{equation}
\sum_{j:(j,i)\in\mathcal{A}^{\mathrm{cand}}} f_{ji}
-
\sum_{k:(i,k)\in\mathcal{A}^{\mathrm{cand}}} f_{ik}
=
y_i,
\qquad \forall i\neq s,
\label{eq:flow_non_source}
\end{equation}
and for the source node,
\begin{equation}
\sum_{k:(s,k)\in\mathcal{A}^{\mathrm{cand}}} f_{sk}
-
\sum_{j:(j,s)\in\mathcal{A}^{\mathrm{cand}}} f_{js}
=
\sum_{i\in\mathcal{N}^{\mathrm{cand}}\setminus\{s\}} y_i.
\label{eq:flow_source}
\end{equation}

The flow is allowed only on activated edges:
\begin{equation}
0 \le f_{ij} \le M z_{ij},
\qquad \forall (i,j)\in\mathcal{A}^{\mathrm{cand}},
\label{eq:flow_capacity}
\end{equation}
where $M$ is a sufficiently large constant, chosen here as $|\mathcal{N}^{\mathrm{cand}}|-1$.

Constraints \eqref{eq:tree_cardinality}--\eqref{eq:flow_capacity} ensure that the selected topology is both source-connected and radial. The resulting feeder graph is denoted by
\begin{equation}
\mathcal{G}^{\mathrm{feed}} = (\mathcal{N}^{\mathrm{feed}}, \mathcal{E}^{\mathrm{feed}}),
\end{equation}
where $\mathcal{N}^{\mathrm{feed}}$ and $\mathcal{E}^{\mathrm{feed}}$ are the selected nodes and edges.

Once topology synthesis is completed, electrical quantities are assigned to obtain the final simulation-ready network:
\begin{equation}
\mathcal{G}^{\mathrm{elec}} =
(
\mathcal{N}^{\mathrm{elec}},
\mathcal{E}^{\mathrm{elec}},
\mathcal{L},
\mathcal{P}
),
\end{equation}
where $\mathcal{L}$ denotes load allocation and $\mathcal{P}$ denotes electrical parameter assignment. The feeder synthesis problem therefore defines the structural core of GeoDistNet, while the subsequent electrical realization step maps the selected graph into a power-flow-ready feeder model.

\section{Feeder Construction and Electrical Modeling}

This section describes how that formulation is instantiated in GeoDistNet, including candidate-graph extraction from GIS data, edge scoring, mixed-integer feeder synthesis, and the subsequent assignment of electrical parameters and loads. The emphasis is on producing feeder models that remain both geographically grounded and directly usable for electrical simulation.

\subsection{Geographic Data Processing and Candidate Graph Extraction}

GeoDistNet starts from publicly accessible geographic data. In the present implementation, road layout is used as the primary spatial scaffold because practical distribution feeders commonly follow road corridors and rights-of-way. Raw geographic layers are first cleaned to remove duplicated segments, disconnected fragments, and minor geometric artifacts that are irrelevant to feeder construction. The cleaned road graph is written as
\begin{equation}
\mathcal{G}^{\mathrm{road}} =
(
\mathcal{N}^{\mathrm{road}},
\mathcal{E}^{\mathrm{road}}
).
\end{equation}

The candidate graph $\mathcal{G}^{\mathrm{cand}}$ is then derived from $\mathcal{G}^{\mathrm{road}}$ through graph simplification and node filtering. This step reduces purely geometric detail while preserving major intersections, branching points, and corridor continuity. The objective is not to reproduce every road segment, but to retain the dominant spatial structure relevant to feeder synthesis.

For each candidate edge, a composite weight is computed to quantify its suitability for feeder construction:
\begin{equation}
w_{ij} =
\lambda_{\mathrm{d}} d_{ij}
+
\lambda_{\mathrm{c}} c_{ij}^{\mathrm{cls}}
+
\lambda_{\mathrm{b}} c_{ij}^{\mathrm{bend}},
\label{eq:edge_score}
\end{equation}
where $d_{ij}$ is the segment length, $c_{ij}^{\mathrm{cls}}$ is a road-class penalty, and $c_{ij}^{\mathrm{bend}}$ is a geometric complexity penalty. The parameters $\lambda_{\mathrm{d}}, \lambda_{\mathrm{c}}, \lambda_{\mathrm{b}}$ balance geometric economy and engineering plausibility. In practice, \eqref{eq:edge_score} provides the concrete edge scoring used in the mixed-integer feeder synthesis problem.

\subsection{Mixed-Integer Feeder Synthesis}

Given the weighted candidate graph, the feeder topology is obtained by solving the mixed-integer problem introduced in Section II. In the current implementation, the model is solved using Gurobi. The solution returns a connected radial subgraph that minimizes the weighted synthesis cost while satisfying the required-node, radiality, and connectivity constraints.

The role of this optimization step is not to recover a utility switching arrangement exactly. Rather, it is to construct a feeder-compatible topology that remains geographically credible and compact enough for subsequent electrical analysis. The resulting graph is denoted by
\begin{equation}
\mathcal{G}^{\mathrm{feed}} =
(
\mathcal{N}^{\mathrm{feed}},
\mathcal{E}^{\mathrm{feed}}
).
\end{equation}

Because the formulation is built on a candidate graph extracted from real map data, the synthesized feeder preserves the dominant spatial hierarchy of the study area. At the same time, the mixed-integer formulation ensures that the selected network remains radial and source-connected, which are essential properties for feeder-level operation and validation.

\subsection{Electrical Parameter Assignment}

Once the feeder topology has been synthesized, each retained node is mapped to an electrical bus and each retained edge is mapped to a line element. Representative line parameters are assigned using class-dependent templates:
\begin{equation}
r_{ij} = \bar{r}_{t_{ij}} d_{ij}, \qquad
x_{ij} = \bar{x}_{t_{ij}} d_{ij},
\end{equation}
where $t_{ij}$ denotes the selected line or road class, and $\bar{r}_{t_{ij}}$, $\bar{x}_{t_{ij}}$ are the corresponding resistance and reactance per unit length.

This assignment does not attempt to reproduce utility-specific conductor inventories. Instead, the adopted parameters are intended to be representative enough for feeder-level engineering studies while remaining transparent and reproducible. The present paper therefore emphasizes electrical usability rather than exact asset calibration.

\subsection{Load Allocation}

After assigning line parameters, loads are distributed across retained buses using a spatial weighting proxy. For each retained node $i$, the weight is defined as
\begin{equation}
\omega_i =
\frac{(A_i+\epsilon)^{\eta}}
{(\delta_i+\epsilon)^{\beta}},
\end{equation}
where $A_i$ is local activity intensity, $\delta_i$ is the distance from node $i$ to the nearest demand centroid, and $\epsilon>0$ is a small regularization parameter. Larger activity intensity and shorter distance both increase the likelihood that a node should carry more demand.

Given a target total feeder demand $P^{\mathrm{tot}}$, the active load assigned to each retained bus is
\begin{equation}
P_i^{\mathrm{load}}
=
P^{\mathrm{tot}}
\frac{\omega_i}{\sum_{k\in\mathcal{N}^{\mathrm{feed}}}\omega_k},
\end{equation}
while the reactive load is computed from an assumed power factor:
\begin{equation}
Q_i^{\mathrm{load}}
=
P_i^{\mathrm{load}}
\tan\!\left(\arccos(\mathrm{pf}_i)\right).
\end{equation}

Through this procedure, the synthesized feeder is endowed with a spatially differentiated load pattern without requiring confidential customer-level utility data. The final feeder model is summarized as
\begin{equation}
\mathcal{M} =
\left\{
\mathcal{N}^{\mathrm{feed}},
\mathcal{E}^{\mathrm{feed}},
P_i^{\mathrm{load}},
Q_i^{\mathrm{load}},
r_{ij},
x_{ij},
V^{\mathrm{slack}}
\right\}.
\end{equation}

\subsection{Export and Validation Quantities}

The synthesized feeder is exported to \texttt{pandapower} for electrical validation. In this work, validation focuses on three basic checks: radiality, voltage deviation, and branch loading. Voltage feasibility is assessed through the maximum deviation from nominal voltage,
\begin{equation}
\Delta V_{\max}
=
\max_{i\in\mathcal{N}^{\mathrm{feed}}}
|V_i - 1|,
\end{equation}
and thermal loading is assessed through the maximum branch loading ratio,
\begin{equation}
\rho_{\max}
=
\max_{(i,j)\in\mathcal{E}^{\mathrm{feed}}}
\frac{|S_{ij}|}{S_{ij}^{\max}}.
\end{equation}

The mixed-integer formulation in Section II defines the optimization core of GeoDistNet. In implementation, this synthesis stage takes the candidate graph, source node, and required-node set as inputs, computes the edge costs, and returns a connected radial feeder graph that satisfies the spatial and structural constraints. Algorithm~\ref{alg:feeder_synthesis} summarizes this procedure. The output of Algorithm is the selected feeder graph $\mathcal{G}^{\mathrm{feed}}$, defined by the retained node and edge sets. Once this graph has been obtained, the remaining steps of GeoDistNet are constructive rather than combinatorial: representative electrical parameters are assigned to retained edges, loads are distributed across retained buses, and the resulting network is exported for load-flow validation. In this way, the synthesis algorithm serves as the structural core of the overall framework.

\begin{algorithm}[!t]
\caption{MIP-Based Feeder Synthesis in GeoDistNet}
\label{alg:feeder_synthesis}
\begin{algorithmic}[1]
\Require Candidate graph $\mathcal{G}^{\mathrm{cand}} = (\mathcal{N}^{\mathrm{cand}}, \mathcal{E}^{\mathrm{cand}})$, source node $s$, required node set $\mathcal{N}^{\mathrm{req}}$
\Ensure Synthesized radial feeder graph $\mathcal{G}^{\mathrm{feed}}$

\State Compute geographic and engineering edge attributes for all $(i,j)\in\mathcal{E}^{\mathrm{cand}}$
\State Form edge weights using \eqref{eq:feeder_obj}
\State Define binary decision variables $y_i$ and $z_{ij}$
\State Enforce node-edge coupling and required-node constraints
\State Impose radiality condition \eqref{eq:tree_cardinality}
\State Construct the directed arc set $\mathcal{A}^{\mathrm{cand}}$ induced from $\mathcal{E}^{\mathrm{cand}}$
\State Introduce flow variables $f_{ij}$ and enforce connectivity constraints \eqref{eq:flow_non_source}--\eqref{eq:flow_capacity}
\State Solve the resulting mixed-integer optimization problem
\State Extract the selected node set $\mathcal{N}^{\mathrm{feed}}$ and edge set $\mathcal{E}^{\mathrm{feed}}$
\State Return $\mathcal{G}^{\mathrm{feed}} = (\mathcal{N}^{\mathrm{feed}}, \mathcal{E}^{\mathrm{feed}})$
\end{algorithmic}
\end{algorithm}

\section{Case Study and Validation}

\subsection{Study Setup}

A residential area in Melbourne, with its street layout extracted from publicly available OpenStreetMap data \cite{map2017open}, is selected as the case-study region to instantiate the proposed workflow. The case is configured to represent 360 households, whose demand is spatially allocated to aggregated load locations in the synthesized feeder model. To define the feeder-level loading baseline, the aggregated demand is parameterized with a peak active load of 1.44 MW. This setting provides a medium-scale residential test case for examining the topology and electrical behavior of the generated distribution feeder without relying on utility-confidential network data.

To assess the electrical response of the synthesized feeder under different demand levels, three loading scenarios are considered. The sanity scenario is introduced as a light-load case for basic consistency checking, the representative scenario is used as the main operating condition for interpreting normal feeder behavior, and the stressed scenario is used to examine network performance under heavier loading. Across all three scenarios, the spatial distribution of demand is kept unchanged, so that the comparison isolates the effect of load level rather than changes in customer placement or feeder structure. The synthesized feeder is then exported to \texttt{pandapower} for steady-state validation.


\begin{table}[!bp]
\caption{Summary of the Generated Feeder}
\label{tab:summary}
\centering
\scriptsize
\setlength{\tabcolsep}{2.5pt}
\begin{tabular}{p{0.38\columnwidth} c p{0.38\columnwidth} c}
\toprule
Metric & Value & Metric & Value \\
\midrule
Number of buses           & 95      & Number of branches          & 94 \\
Total line length (km)         & 9.68  & Peak active load (MW)          & 1.44  \\
Number of load points     & 82      & Total households            & 360 \\
Feeder depth (hops)              & 22      & Root branches               & 3 \\
Mean depth (hops)               & 11.20   & Number of leaves            & 34 \\
Number of branching nodes & 30      & Mean line length (km)           & 0.103 \\
Max line length (km)         & 0.398  & Mean households per load point & 4.39 \\
\bottomrule
\end{tabular}
\end{table}

\subsection{Generated Network Characteristics}

Table~\ref{tab:summary} summarizes the main structural characteristics of the generated feeder, including network size, route length, household representation, and demand aggregation. The synthesized network contains 95 buses and 94 branches, indicating a nontrivial radial feeder rather than a small illustrative graph. The case represents 360 households aggregated into 82 load points, so demand is modeled at a feeder-relevant level of aggregation rather than at the level of individual customer connections. This reflects the intended use of the case setup, in which multiple households are mapped to shared load buses while preserving the overall spatial demand pattern.

The structural metrics in Table~\ref{tab:summary} also indicate a clear downstream hierarchy. A feeder depth of 22 hops and a mean depth of 11.20 hops show that the synthesized topology is neither a shallow star-shaped structure nor a trivial chain. In addition, the presence of 30 branching nodes and 34 leaves suggests a feeder organized around a limited number of upstream corridors with multiple downstream laterals. This is consistent with the expected organization of a residential distribution feeder, where several primary sections collect demand from a broader set of terminal branches.

From a geometric perspective, the total route length is 9.68~km, with a mean line length of 0.103~km and a maximum segment length of 0.398~km. These values indicate that the synthesis process suppresses excessive short-segment fragmentation while retaining the larger-scale form of the study area. The resulting network therefore lies between the raw street graph and a highly compressed benchmark-style abstraction: the dominant spatial constraints of the underlying area are retained, while the final topology remains compact enough for feeder-level electrical analysis.

\begin{figure}[!t]
\centering
\includegraphics[width=0.85\columnwidth]{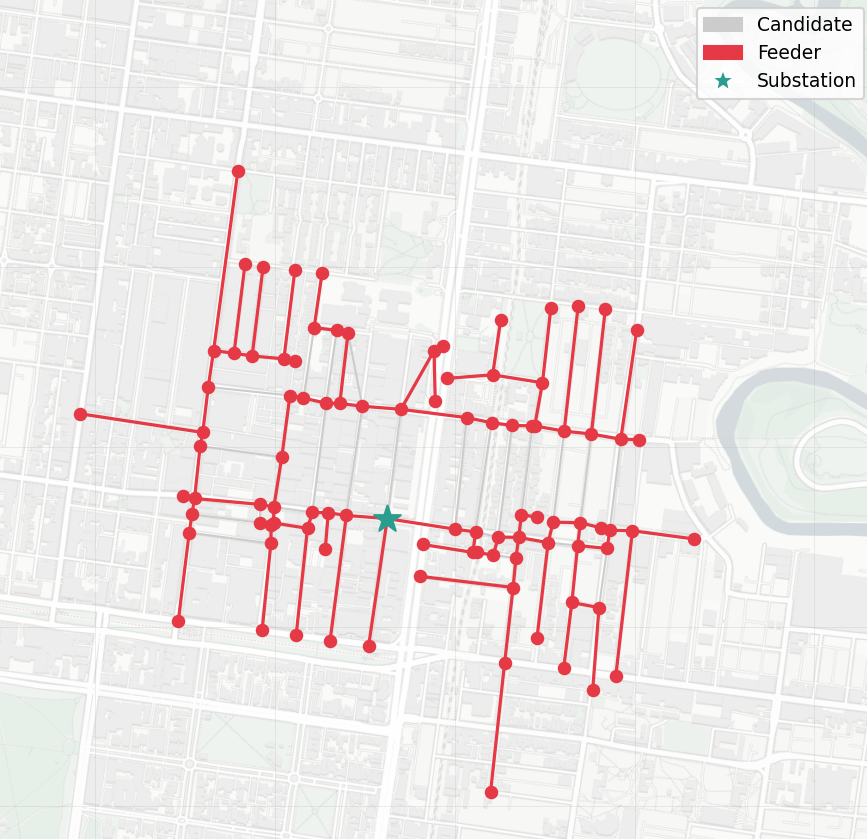}
\caption{Study area and synthesized feeder topology. The grey layer shows the underlying street-layout geometry of the selected Melbourne study area, while the red network denotes the radial feeder synthesized by GeoDistNet and the cyan marker indicates the source bus.}
\label{fig:overlay}
\end{figure}

This characteristic is visible in Fig.~\ref{fig:overlay}. The synthesized feeder does not reproduce the full street graph directly. Instead, it follows the major spatial corridors of the study area and selectively extends secondary branches toward distributed demand locations. The source is positioned near the central part of the network, from which several primary corridors extend outward before splitting into shorter laterals. As a result, the generated topology remains geographically interpretable while avoiding the excessive local detail that would arise if the original geographic graph were used directly for simulation. Fig.~\ref{fig:overlay} therefore illustrates not only the final feeder layout, but also the level of abstraction achieved by the proposed workflow: the network remains spatially grounded, yet is simplified to a form suitable for steady-state feeder studies.

The aggregation level reported in Table is also relevant from a modeling perspective. Although the case represents 360 households, these are mapped to 82 load points rather than modeled as individual terminal connections. This means that the synthesized feeder captures feeder-level demand geography without introducing excessive customer-level detail. In practical terms, the resulting representation is detailed enough to preserve nonuniform downstream demand pickup, yet compact enough to remain computationally manageable for repeated simulation studies.

\subsection{Power-Flow Validation}

\begin{figure}[!tp]
\centering
\includegraphics[width=1\columnwidth]{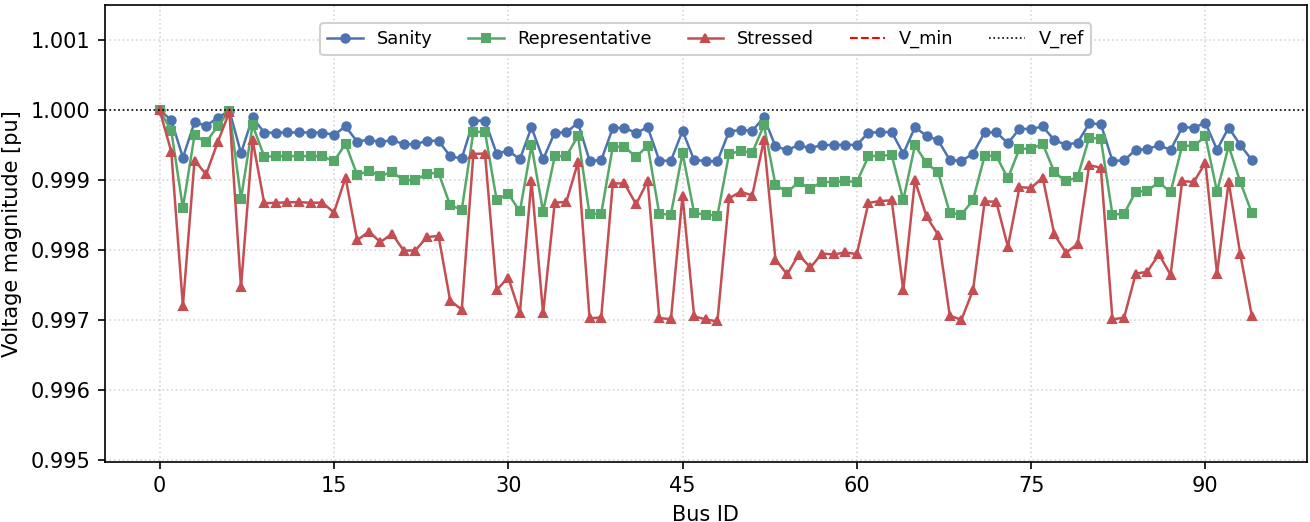}
\caption{Illustrative load-flow validation results for the synthesized feeder under the sanity, representative, and stressed loading scenarios.}
\label{fig:pf}
\end{figure}

The synthesized feeder is further examined through steady-state power-flow analysis in \texttt{pandapower}. Fig.~\ref{fig:pf} plots the bus-voltage magnitudes under the three loading levels defined in Subsection~A.

The effect of loading level is immediately visible. The sanity case remains closest to the nominal reference, the representative case lies below it, and the stressed case forms the lowest profile across almost the entire feeder. This ordered separation is expected, but it is still useful: it shows that the generated feeder responds to increasing demand in a stable and physically interpretable way, rather than producing irregular or contradictory voltage behavior.

The voltage curves also show localized drops and partial recoveries along the bus index, especially in the stressed case. This pattern is consistent with a feeder that contains multiple downstream branches of unequal depth and loading, rather than a highly regularized or trivial topology. In other words, the synthesized network behaves like a structured radial feeder with spatially differentiated demand accumulation.

At the same time, all voltages remain above the lower bound \(V_{\min}\), including under the stressed loading condition. For the present paper, this is sufficient to demonstrate that the generated feeder is not only geographically grounded, but also usable as a simulation-ready model for steady-state feeder studies.

\section{Conclusion}

This paper presented GeoDistNet, an open-source workflow for constructing synthetic distribution feeders from publicly available geographic information. Starting from map-derived street-layout data, the proposed framework builds a candidate graph, synthesizes a feeder-compatible radial topology through mixed-integer optimization, assigns representative electrical parameters and aggregated demand, and exports the resulting network to a simulation-ready model for steady-state analysis.

The Melbourne case study showed that the generated feeder remains geographically interpretable, topologically structured, and directly usable in \texttt{pandapower} under multiple loading levels. The synthesized network preserves the dominant spatial organization of the study area while avoiding the excessive geometric detail of the raw street graph. The resulting feeder therefore occupies a practically useful intermediate level of abstraction between geographic data and electrical network simulation.

The study also demonstrated that GeoDistNet can support different levels of geographic fidelity, ranging from structured GIS-based synthetic cases to real-world OpenStreetMap-derived study areas, while maintaining a common downstream validation workflow. This makes the framework suitable not only for feeder generation itself, but also for reproducible benchmarking, comparative topology studies, and future distribution-level analysis when detailed utility feeder data are unavailable.

Future work will extend the present framework in three directions. First, the electrical parameterization and load allocation stages can be refined using richer regional data and more detailed customer representations. Second, time-varying demand and distributed energy resources can be incorporated to support quasi-static and time-series studies. Third, larger and more diverse real-world case studies can be used to further examine the robustness and transferability of the proposed workflow across different urban and suburban settings.

\bibliographystyle{IEEEtran}
\bibliography{references}

@article{Baran1989,
  author  = {M. E. Baran and F. F. Wu},
  title   = {Network Reconfiguration in Distribution Systems for Loss Reduction and Load Balancing},
  journal = {IEEE Transactions on Power Delivery},
  volume  = {4},
  number  = {2},
  pages   = {1401--1407},
  year    = {1989}
}

@article{Dolatabadi2020,
  author  = {Sarineh Hacopian Dolatabadi and Maedeh Ghorbanian and Pierluigi Siano and Nikos D. Hatziargyriou},
  title   = {An Enhanced IEEE 33 Bus Benchmark Test System for Distribution System Studies},
  journal = {IEEE Transactions on Power Systems},
  volume  = {36},
  number  = {3},
  pages   = {2565--2572},
  year    = {2021}

}

@article{Postigo2017,
  author  = {Fernando E. Postigo Marcos and Carlos Mateo Domingo and Tom{\'a}s G{\'o}mez San Rom{\'a}n and Bryan S. Palmintier and Bri-Mathias Hodge and Venkat Krishnan and Fernando de Cuadra Garc{\'i}a and Barry A. Mather},
  title   = {A Review of Power Distribution Test Feeders in the United States and the Need for Synthetic Representative Networks},
  journal = {Energies},
  volume  = {10},
  number  = {11},
  pages   = {1896},
  year    = {2017}

}

@article{Birchfield2017,
  author  = {Adam B. Birchfield and Ti Xu and Kathleen M. Gegner and Komal S. Shetye and Thomas J. Overbye},
  title   = {Grid Structural Characteristics as Validation Criteria for Synthetic Networks},
  journal = {IEEE Transactions on Power Systems},
  volume  = {32},
  number  = {4},
  pages   = {3258--3265},
  year    = {2017}
}

@article{Mateo2020,
  author  = {Carlos Mateo and Fernando Postigo and Fernando de Cuadra and Tom{\'a}s G{\'o}mez San Rom{\'a}n and Tarek Elgindy and Pablo Due{\~n}as and Bri-Mathias Hodge and Venkat Krishnan and Bryan Palmintier},
  title   = {Building Large-Scale U.S. Synthetic Electric Distribution System Models},
  journal = {IEEE Transactions on Smart Grid},
  volume  = {11},
  number  = {6},
  pages   = {5301--5313},
  year    = {2020}

}

@article{Liang2021,
  author  = {Ming Liang and Yao Meng and Jiyu Wang and David L. Lubkeman and Ning Lu},
  title   = {FeederGAN: Synthetic Feeder Generation via Deep Graph Adversarial Nets},
  journal = {IEEE Transactions on Smart Grid},
  volume  = {12},
  number  = {2},
  pages   = {1163--1173},
  year    = {2021},
  doi     = {10.1109/TSG.2020.3025259}
}

@inproceedings{Saha2019,
  author    = {Shammya Shananda Saha and Eran Schweitzer and Anna Scaglione and Nathan G. Johnson},
  title     = {A Framework for Generating Synthetic Distribution Feeders using OpenStreetMap},
  booktitle = {2019 North American Power Symposium (NAPS)},
  year      = {2019}
}

@inproceedings{Trpovski2018,
  author    = {Andrej Trpovski and Dante Fernando Recalde Melo and Thomas Hamacher},
  title     = {Synthetic Distribution Grid Generation Using Power System Planning: Case Study of Singapore},
  booktitle = {2018 53rd International Universities Power Engineering Conference (UPEC)},
  year      = {2018}
}

@article{Thurner2018,
  author  = {Leon Thurner and Alexander Scheidler and Florian Sch{\"a}fer and Jan-Hendrik Menke and Julian Dollichon and Friederike Meier and Steffen Meinecke and Martin Braun},
  title   = {pandapower---An Open-Source Python Tool for Convenient Modeling, Analysis, and Optimization of Electric Power Systems},
  journal = {IEEE Transactions on Power Systems},
  volume  = {33},
  number  = {6},
  pages   = {6510--6521},
  year    = {2018}
}

@article{Boeing2017,
  author  = {Geoff Boeing},
  title   = {OSMnx: New Methods for Acquiring, Constructing, Analyzing, and Visualizing Complex Street Networks},
  journal = {Computers, Environment and Urban Systems},
  volume  = {65},
  pages   = {126--139},
  year    = {2017}
}

@inproceedings{Dugan2011,
  author    = {Roger C. Dugan and Thomas E. McDermott},
  title     = {An Open Source Platform for Collaborating on Smart Grid Research},
  booktitle = {2011 IEEE Power and Energy Society General Meeting},
  year      = {2011}
}

@article{map2017open,
  title={Open street map},
  author={Map, Open Street},
  journal={Acessado em},
  volume={12},
  year={2017}
}

\end{document}